# Dynamics of Disagreement: Large-Scale Temporal Network Analysis Reveals Negative Interactions in Online Collaboration

Milena Tsvetkova [a], Ruth García-Gavilanes [a], and Taha Yasseri [a,*]

[a] Oxford Internet Institute, University of Oxford, Oxford OX1 3JS, United Kingdom

[*] To whom correspondence should be addressed. Address: Oxford Internet Institute, 1 St Giles, Oxford OX1 3JS, United Kingdom. Telephone: +44 1865 287 229. E-mail: taha.yasseri@oii.ox.ac.uk

**Abstract**

Disagreement and conflict are a fact of social life. However, negative interactions are rarely explicitly declared and recorded and this makes them hard for scientists to study. In an attempt to understand the structural and temporal features of negative interactions in the community, we use complex network methods to analyze patterns in the timing and configuration of reverts of article edits to Wikipedia. We investigate how often and how fast pairs of reverts occur compared to a null model in order to control for patterns that are natural to the content production or are due to the internal rules of Wikipedia. Our results suggest that Wikipedia editors systematically revert the same person, revert back their reverter, and come to defend a reverted editor. We further relate these interactions to the status of the involved editors. Even though the individual reverts might not necessarily be negative social interactions, our analysis points to the existence of certain patterns of negative social dynamics within the community of editors. Some of these patterns have not been previously explored and carry implications for the knowledge collection practice conducted on Wikipedia. Our method can be applied to other large-scale temporal collaboration networks to identify the existence of negative social interactions and other social processes.

**Introduction**

In the last few decades, network science has significantly advanced our understanding of the structure and dynamics of the human social fabric (*1*). Much of the research, however, has focused on positive relations and interactions such as friendship and collaboration (*2*, *3*). Considerably less is known about networks of negative social interactions such as distrust, disapproval, and disagreement. Although such interactions are not as common, they strongly affect people's psychological well-being, physical health, and work performance (*4–8*).

Previous work has investigated social networks with negative relations mainly from the perspective of structural balance theory (*9–11*). According to this theory, the structure of social networks with both positive and negative relations, also known as "signed networks," should follow principles such as "the friend of my friend is my friend" and "the enemy of my friend is my enemy." Empirical research has found evidence that balanced triads are indeed dominant (*12–14*), that the global network structure tends to be balanced (*15*, *16*), and that networks evolve to increase their balance (*17*). While this research illuminates how negative social relations interact with positive social relations, we still know very little about the structure and dynamics of negative interactions. For example, we have limited knowledge regarding the extent to which interaction mechanisms that we have observed for positive interactions, such as direct reciprocity and generalized, or "pay-it-forward" reciprocity, transfer to negative interactions.

Motivated by these gaps in our knowledge, we study a large online collaboration community and analyze the timing and configuration of sequences of contributions to identify patterns of negative social interactions among users. In such communities, users sometimes undo or downrate contributions made by other users. Most often these actions intend to maintain and improve the collaborative project. However, it is also possible that these actions are social in nature. Although we cannot tell whether a particular content-related action is dictated by hostility towards another user, we can study the dynamic patterns of such actions in aggregate to find if social processes occur in the community and if so, the form they take. This is what we do here.

We focus on six interactions, each of which is a sequence of two content-related actions between two or three users. We analyze the six interactions as temporal "motifs" (*18*, *19*). We compare the observed frequency and dynamics of the motifs to those generated by a null model without any systematic clustering of actions in time for individuals but with equivalent daily patterns and community structure. In addition, we study the status effects associated with the motifs, where we use the contributor's volume of activity on the collaborative platform as a proxy for status. We confirm that the interaction is social if the corresponding motif exhibits time patterns that significantly differ from those observed in the null model and if the status effects fit with our expectations based on social psychology and results from previous studies. To present a richer picture of the observed patterns of interaction, we also look for prominent cultural effects associated with the motifs.

Our data come from Wikipedia, the largest free-access online encyclopedia. There has been much research on the coverage, quality, and biases of information provided by Wikipedia (*20–22*) and the controversy of topics and articles on it (*23*, *24*). Wikipedia and its community of editors have also been extensively studied to investigate fundamental social processes. For example, using the Wikipedia community as a test case, previous research has shown that social rewards improve individual effort (*25*), language choice reveals power relations (*26*), success breeds success (*27*), and conflict can be both productive and counter-productive (*28–30*).

Wikipedia is edited by millions of volunteers. Usually, editors add text to articles, polish the existing text, or correct minor errors. Sometimes, however, they revert other editors – they undo other editors' contributions by restoring an earlier version of the article. The option to "revert" an edit is a specific feature of the Wikimedia software that was originally implemented to cope with vandalism. Reverts are technically easy to detect regardless of the context and the language and hence, they enable analysis at the scale of the whole system.

Although reverts on Wikipedia are intended to improve the encyclopedia's content, previous research has acknowledged that they could also imply negative social interactions. According to one perspective, reverts constitute maintenance work and imply conflict and coordination costs (*31*, *32*). According to another perspective, the number of words added, deleted, and restored can be used to visualize the bipolarity of the network of editors behind an article, which in turn is associated with the perceived controversy of the article (*33*). In fact, information on reverts can be used to construct a measure of article controversy and study editorial wars in relation to topics and articles (*34–36*). Further, the time-series of reverts and non-revert edits can be analyzed to infer states of conflict and cooperation over the history of an article (*37*). Reverts can also be combined with other actions with negative valence on Wikipedia, such as voting against an editor's promotion to admin level, to infer a signed network of relations (*14*). Finally, Wikipedia reverts have been used to develop formal mathematical models of opinion dynamics explaining conflict in collaborative environments in general and beyond the Wikipedia platform (*38*, *39*).

Similarly to these previous studies, we acknowledge that a revert is not necessarily socially motivated but reverts consistently occurring in certain patterns could be signs of negative social interactions. Going beyond most existing research, we analyze the dynamics of the entire networks of reverts by considering various temporal motifs. Previous work has analyzed the distribution and structure of editing sequences and editing motifs on Wikipedia (*40*, *41*) and demonstrated the potential of this method to reveal unknown patterns in online collaboration. We here apply this method to reverts only in order to reveal patterns of negative interactions on Wikipedia. We are also the first to analyze data of reverts on multiple language editions of Wikipedia as social networks. The networks we analyze comprise about 4.7 million reverts over the period January 2001 – October 2011 among the editors of 13 different language editions of Wikipedia: English, Spanish, German, Japanese, French, Portuguese, Chinese, Hebrew, Arabic, Hungarian, Persian, Czech, and Romanian. The evidence we find suggests that certain social processes interfere with how knowledge is negotiated on Wikipedia. The approach we assume can be applied to other online collaboration networks and content communities to identify the presence of similar interferences.

**Results**

To build the revert networks, we create time-stamped directed links going from the reverters to the reverted users. The aggregate revert networks in different language editions vary in size from 4,019 to 701,169 nodes (Table 1). All networks have average clustering and reciprocity that are significantly larger than expected in a simple random graph with the same size and density (these results were confirmed in exponential-random graph models (*42*, *43*) not reported here). The giant components in the networks have short path lengths in the range 3.04–4.16. The out-degree and in-degree distributions are fat-tailed, consisting of a few editors with many reverts and many editors with only a few reverts (Supplementary Fig. S1). Further, the out-degree assortativity in the directed networks is significantly negative, implying that users who revert frequently tend to revert users who revert rarely. There is no evidence for in-degree assortativity.

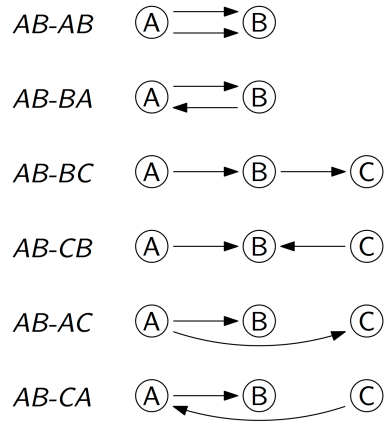

**Fig. 1.** The six two-event temporal motifs studied. An arrow from *A* to *B* signifies that *A* reverted the content contributed by *B*. An arrow on top or to the left indicates an earlier event than an arrow on the bottom or to the right.

**Table 1. Descriptive statistics for the 13 Wikipedia revert networks studied.**

| | Directed, multi-link | | | Directed, single-link | | | Undirected |
|---|---|---|---|---|---|---|---|
| Language | Num. nodes | Num. links | Reciprocity | Out-degree assortativity | In-degree assortativity | Avg. path length | Avg. clustering |
| English | 701169 | 3516109 | 0.183 | -0.082 | 0.031 | 4.01 | 0.055 |
| Spanish | 58365 | 265291 | 0.202 | -0.096 | -0.074 | 3.49 | 0.149 |
| German | 38334 | 151414 | 0.224 | -0.074 | 0.054 | 4.08 | 0.034 |
| Japanese | 32314 | 150265 | 0.230 | -0.097 | -0.073 | 4.16 | 0.042 |
| French | 31727 | 122374 | 0.191 | -0.126 | 0.026 | 3.92 | 0.051 |
| Portuguese | 31287 | 159963 | 0.226 | -0.117 | -0.042 | 3.32 | 0.199 |
| Chinese | 19359 | 114459 | 0.260 | -0.105 | -0.099 | 3.29 | 0.145 |
| Hebrew | 11778 | 79989 | 0.264 | -0.089 | -0.047 | 3.11 | 0.283 |
| Arabic | 7365 | 28619 | 0.186 | -0.161 | -0.047 | 3.24 | 0.197 |
| Hungarian | 6114 | 23657 | 0.191 | -0.154 | 0.026 | 3.38 | 0.126 |
| Persian | 6013 | 34879 | 0.294 | -0.164 | 0.160 | 3.10 | 0.264 |
| Czech | 4542 | 19498 | 0.239 | -0.103 | -0.030 | 3.27 | 0.160 |
| Romanian | 4019 | 12974 | 0.187 | -0.186 | -0.209 | 3.04 | 0.179 |

Our analysis focuses on temporal motifs, which are classes of event sequences that are similar not only in the topology but also in the temporal order of the events (*44*). In particular, we analyze the two-event temporal motifs in which after *A* reverts *B*, *A* reverts *B* again (*AB-AB*), *B* reverts *A* back (*AB-BA*), *B* reverts *C* (*AB-BC*), *C* reverts *B* (*AB-CB*), *A* reverts *C* (*AB-AC*), and *C* reverts *A* (*AB-CA*), as illustrated in Figure 1. To identify the motifs, we look at every revert and identify if and when a response of these six forms occurs, restricted to a time window of 24 hours. The response may happen either in the same or in a different article. Further, the response may occur immediately after the original revert or alternatively, the reverter and the reverted may be involved in other reverts in-between the original revert and the response.

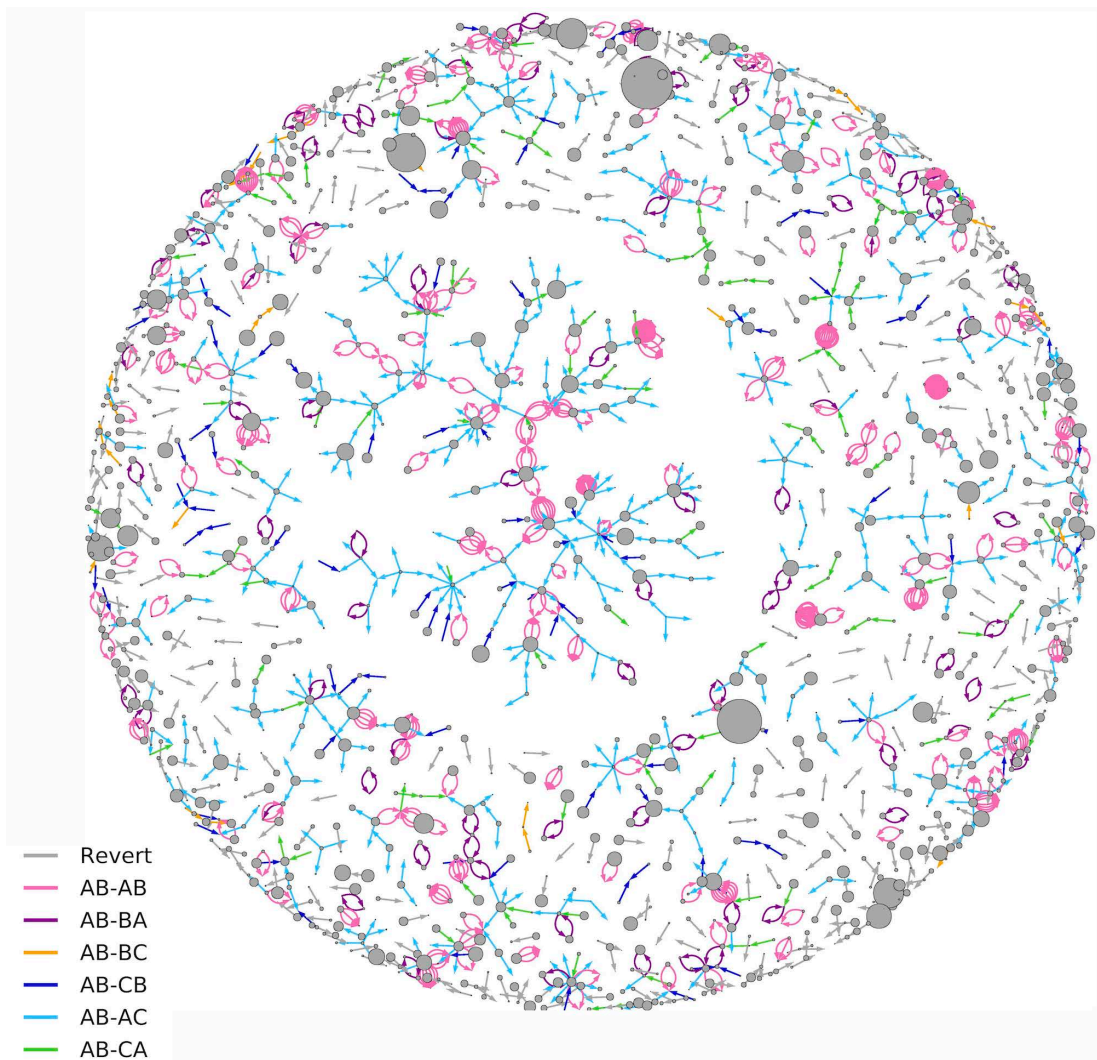

**Fig. 2.** An example network of all reverts done in the English language Wikipedia within one day (January 15, 2010). The six temporal motifs that we investigate are color-coded. A link points from the reverter to the reverted user. The area of the nodes corresponds to the total number of edits by the user. A revert can be counted as part of more than one motif but for presentation purposes, we show only one, with the following priority: *AB-AB*, *AB-BA*, *AB-AC*, *AB-CA*, *AB-CB*, *AB-BC*. The network is drawn using the Kamada-Kawai force-directed algorithm.

Figure 2 illustrates the identified motifs in an example time-snippet of the English-language network. If the actions in the temporal motif are socially motivated, then the motif represents a particular social interaction. For example, if *B* reciprocates a revert by *A* (*AB-BA*), this could indicate "self-defense" (undoing the "attack") or "retaliation" (responding to the "attacker" with the same); if *B* reverts someone else after being reverted (*AB-BC*) then this could be a case of "pay it forward" retaliation. Our method does not allow us to establish the exact meaning of the social interaction as it is extremely hard to make causal claims with observational data. What we instead do is to look for evidence that the interaction is social. We assume if this is the case, the temporal motif will be common, occur fast, be associated with social status effects, and possibly, be related to prominent cultural differences between editors working in different languages. Thus, our analyses test against the null hypothesis that the six motifs we observe are not socially motivated and that they do not occur more often than random chance and that they are not associated with status effects as predicted by social psychology theories and previous research.

Thus, to establish evidence for social interactions, we need appropriate null models for comparison. First, if the motifs occur intentionally and systematically, they should exhibit time patterns that differ from any random collection of pairs of reverts. Typically, in motif analysis, the null model is a suitably randomized version of the empirical data (*18*). The null model we use preserves the network structure and major temporal features such as burstiness but shuffles the timestamps of events within individuals and within a window of 24 hours. The shuffle occurs with these restrictions to preserve the general patterns of individual activity and to identify each social interaction against this expected sequence of events.

Second, if the motifs represent social interactions, they should exhibit social status patterns that differ from those associated with any other pairs of reverts. We rely on common social psychology theories and previous research to derive expected status effects for most of the interactions we study and we test them below. To measure status, we use the base-ten logarithm of the number of edits the editor has completed by the time of the revert under question. This measure has a number of advantages to alternative operationalizations of status in the community of editors. The number of edits gives us a well-balanced continuous measure of experience and seniority that is easy to compare across different language versions of Wikipedia and that can be used over the entire ten-year period of observation. Importantly, the number of edits has been shown to be the strongest predictor for promotion to administrator status (*45*).

Here, we visualize results only for the largest network we study, the English language Wikipedia; the results for the other 12 networks are reported in the *Supplementary Information*. Since we test multiple hypotheses on multiple large networks, we approach the significance of results with more scrutiny. We take a conservative approach and only interpret the results that are consistent in direction and significance across different languages and operationalizations (see *Materials and Methods*). Most of the differences observed by language are not robust to different operationalizations and hence, apart from the few cases mentioned below, we do not attribute different results by language to cultural differences.

***AB-AB.*** In our data, the *AB-AB* motif is more common and with a shorter time interval between the two actions than expected (Fig. 3A; see also Supplementary Fig. S2, Supplementary Table S1, and Supplementary Table S2). In general, more active and experienced editors do more reverts and hence, the reverters tend to have higher status than the reverted user. The same holds for the *AB-AB* motif, without any significant deviations for most languages (see Supplementary Fig. S8 and Supplementary Table S3).

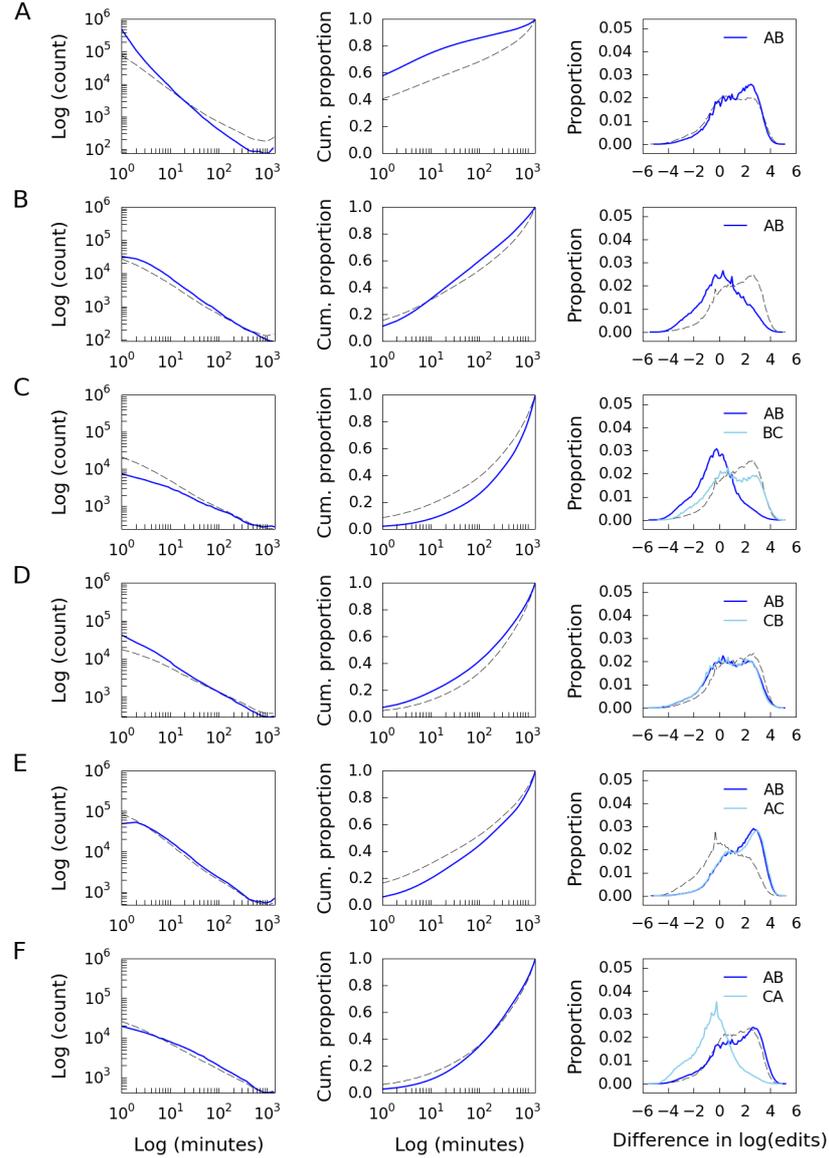

**Fig. 3.** The frequency, rate, and status difference for the six temporal motifs. Results are shown for the English language version of Wikipedia. The dashed line shows the mean expectation from the null model (first two columns; error bars are too small to be shown) or all other reverts (third column). The results on frequency are quantified using z-scores for counts. The results on rate are quantified using multiple measures: Kolmogorov-Smirnov test statistic, mean time, and skewness. The results on status are quantified using regression analyses with standard errors adjusted for clustering within reverter and reverted. (A) The *AB-AB* temporal motif occurs more commonly and with a faster response rate than expected. The difference in status between the reverter and the reverted user is significantly larger for English but this result is not significant for the majority of languages. (B) The *AB-BA* motif is more common and occurs at a faster rate than expected. The results are consistent for all languages except Japanese and Chinese. Regardless of language, the editors involved have significantly smaller status differences and in fact, on average, *B* has smaller status than *A*. (C) The *AB-BC* motif does not occur more commonly and has slower response rate than expected. The difference in status between *A* and *B* is significantly smaller and in fact, negative. (D) The *AB-CB* motif is not more common than expected but usually occurs at a faster rate. There is no evidence for status effects, as the results observed here for English Wikipedia are not consistent across the other languages. (E) The *AB-AC* motif does not occur more commonly and is in fact slower than expected. The difference in status between the reverter and the reverted user is significantly larger than expected. (F) The *AB-CA* motif occurs more commonly than expected. It is significantly faster for the English Wikipedia here but this result is neither consistent among the other languages nor robust to different measures of the shape of the distribution. The status difference between *C* and *A* is significantly smaller than expected.

This result is somewhat surprising because we expected that individuals that are closer in status and thus, in a sense, status rivals, will be more likely to revert each other repeatedly (*46*).

**AB-BA.** The *AB-BA* motif occurs more commonly and the response rate (the time lapse between *AB* and *BA*) is shorter than what the null model generates (Fig. 3B). However, for the Chinese and Japanese languages the opposite is true – the *AB-BA* motif is not more common and the response time is significantly longer than expected (Supplementary Fig. S3). The revert networks for these two languages do not differ in any other structural aspect from the networks of the other editions. Further, no such deviations are observed in any of the other interactions that we study. Since there is no specific editorial rule that may drive these deviations, we attribute them to cultural differences.

Social comparison theory states that people strive to gain accurate self-evaluations and as a result, they tend to compare themselves to those who are similar (*47*). But since focus on relative performance heightens feelings of competitiveness, rivalry is stronger among similar individuals (*48*). Indeed, we find that for all languages, the users involved in the *AB-BA* motif tend to be closer in status than expected (Supplementary Fig. S9). In fact, the average difference in status is negative, meaning that *A* tends to have lower status than *B*. Additional analyses show that the status effect is driven by both equal-status and low-status reverters (Fig. 4).

We also found that the motif is more likely to occur in the same article compared to a baseline built by randomly pairing each node's reverts (Supplementary Table S4). This suggests that the interaction is largely driven by "self-defense", whereby when editor *A* reverts *B*, *B* reverts that revert. Nevertheless, up to 50% of the back-reverts occur in different articles.

**AB-BC.** The *AB-BC* motif does not occur more commonly than expected and when it happens, the response time is longer (Fig. 3C). The literature suggests that individuals pay it forward to lower-status individuals (*49*, *50*) but this is not the case here. The only status effect we observe is that *B* is likely to revert someone else if *B* was reverted by a lower-status user (Supplementary Fig. S10 and Supplementary Table S3).

**AB-CB.** The *AB-CB* motif is not more common (Fig. 3D) although when it happens, it appears slightly faster (except for the Arabic, Chinese, and Japanese Wikipedias; Supplementary Table S2). In the *AB-CB* motif, *A* and *C* have a leader-follower relationship (*51*) and from organizational theory we know that rational actors imitate each other when they want to maintain relative competitive position (*52*). We thus expected that *C* is likely to be competing for status with *A*, implying that *C* should have slightly lower status than *A*. However, this does not appear to be the case (Supplementary Fig. S11 and Supplementary Table S3). In general, we find no significant status effects associated with the *AB-CB* motif.

**AB-AC.** The *AB-AC* motif is rare in our data, and the time interval between two successive actions is longer than expected (Fig. 3E). The reverts in this motif are executed by users with high status on users with low status (Supplementary Fig. S12). Low-status users do not participate in such reverts (Fig. 4C). We expected these status effects as we know that more senior Wikipedia editors tend to scout out for vandals and watch out for edits made by newcomers. Thus, we do not find evidence for systematic serial attacks but rather for senior editors "keeping the streets clean."

***AB-CA.*** The *AB-CA* motif is more common than expected, although with a response rate as fast as expected. (Fig. 3F). On average, *C* is significantly lower in status than both *A* and *B* (Supplementary Fig. S13). One interpretation of the *AB-CA* motif is that when a revert occurs, a third party intervenes and questions the authority of the reverter. A priori, we expected that an editor of higher status would be more likely to contest another editor's opinion. In contrast, the status effects we observe suggest that "third-party defense" happens due to daring newcomers, rather than experienced and established users.

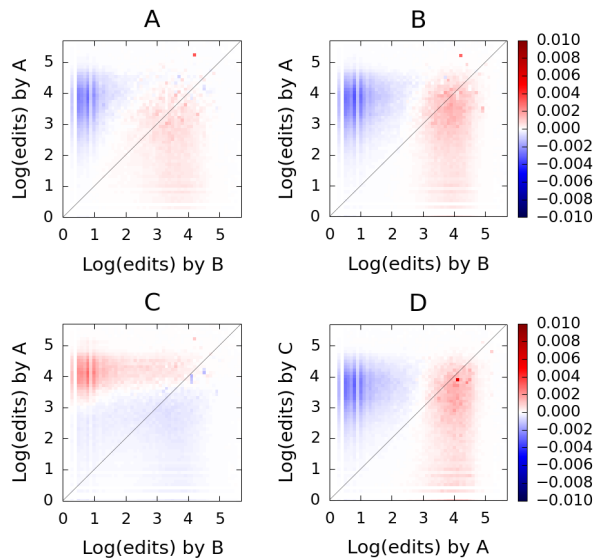

**Fig. 4.** Difference between the observed and expected distribution of status in four of the six temporal motifs. (A) For the *AB-BA* motif, *A* is more likely to be of equal or lower status than *B* compared to all other reverts. (B) For the *AB-BC* motif, *A* is more likely to be of equal or lower status than *B*. (C) For the *AB-AC* motif, *A* is more likely to have high status. (D) For the *AB-CA* motif, *C* is more likely to be of equal or lower in status than *A*.

## Discussion

Large-scale collaboration by volunteers online provides much of the information we obtain and the software products we use today. The repeated interactions of these volunteers have also given rise to communities with shared identity and practice. The social interactions in these communities can in turn induce biases and subjectivities into collaborative public goods. Identifying the social interactions that play a role in these communities is thus an important first step towards understanding the biases in the content and products we consume.

In online collaboration communities, individuals may sometimes negatively evaluate or even undo other individuals' contributions. These actions may be related to the collaborative project but may also indicate negative social interactions among individuals. To better understand the latter, we here proposed to investigate the temporal and structural patterns of sequences of content-related actions.

To exemplify this approach, we analyzed six temporal motifs in the network of reverts among Wikipedia editors. We found evidence that Wikipedia editors systematically revert the same person, revert back their reverter, and come to defend a reverted editor. We did not find evidence that they "pay forward" a revert, coordinate with others to revert an editor, or revert different editors serially. In addition, we found that reciprocation of reverts tends to occur between status equals, while those who revert the reverters of others tend to be with low status. Further, high-status contributors are more likely to be involved in serial reverts. In essence, if senior Wikipedia editors revert, they may continue and revert someone else, they may get reverted by a low status contributor, or they may get reverted back if the editor they reverted is of equal standing.

Some of our findings are intuitive in the context of the structure and organization of Wikipedia. For example, vigilantism is common on Wikipedia and in particular, senior editors are always on the lookout for potential vandals or clueless newcomers. Hence, it is not surprising that serial reverts are not systematic but mainly conducted by senior editors. There are also explicit Wikipedia rules that discourage serial reverts; in particular, the "three-revert rule" prohibits editors from performing more than three reverts on the same page within 24 hours.

The context of Wikipedia also provides alternative explanations for some of our findings. For example, the result that "third-party defenders" tend to be of low status can be due to sock-puppetry, which refers to the practice of maintaining multiple accounts on Wikipedia with the aim to disrupt discussions or avoid editorial restrictions. If this were the case, senior editors would systematically have secondary accounts which they would use to revert the reverts of the edits they have done from their primary account. Such cases undoubtedly occur in our data. Nevertheless, a lot of this secretive behavior could also be happening through anonymous accounts and our analyses excluded those. Future research should investigate whether sock-puppetry indeed dictates the patterns we observed.

Still, some of our findings are unexpected. For example, although editorial wars and back-and-forth conflict have been extensively studied before, we were surprised to find that reciprocation of reverts is common except among the editors of the Japanese and Chinese editions of Wikipedia. The Japanese and Chinese cultures are known as "honor-shame cultures" so one plausible explanation is that editors in these languages avoid direct conflict in fear of ostracism (*53*, *54*). In general, we treated most differences by language as a robustness test, as we acknowledge that multiple hypothesis testing on large amounts of data is bound to produce significant results in either direction (*55*). Nevertheless, it is worthwhile to use more robust techniques to specifically investigate any cultural differences between the different language communities on Wikipedia.

Overall, our aim was to quantify the extent to which certain types of negative social interactions play a role in online collaborative communities, using the particular example of Wikipedia. We focused our analysis on the macro-level and thus, admittedly, we miss the nuanced understanding that thick ethnographic descriptions can produce (*56*). Nevertheless, our findings provide the foundation and hopefully, an inspiration, for such more focused studies in the future. Our finding that certain types of negative social interactions and status considerations interfere with knowledge production on Wikipedia has practical implications both for controlling the quality of content and for maintaining editors' productive involvement on Wikipedia. Future research should use more in-depth analyses at a smaller scale to verify and qualify our findings.

Although our study revealed new knowledge about the structure and dynamics of negative interactions on Wikipedia, our findings do not automatically extend to other social systems. Still, in alignment with Keegan et al. (*41*), we strongly believe in the research potential of sequence and temporal motif analysis for illuminating this research problem. Thus, future research should apply our approach to open-source software projects and bulletin-board-type systems to examine the extent to which our findings generalize to other collaborative communities online.

**Methods**

**Wikipedia data.** Our data contain who reverts whom, when, and in what article. To obtain this information, we analyzed the Wikipedia XML Dumps (https://dumps.wikimedia.org/mirrors.html) of the 13 language editions we study. To detect restored versions of an article, a hash was calculated for the complete article text following each revision and the hashes were compared between revisions. To create the network, we assumed that a link goes from the editor who restored an earlier version of the article (the "reverter") to the editor who made the revision immediately after that version (the "reverted"). In an alternative operationalization, we could have created links from the reverter to every editor who made a revision in the time between the restored version and the revert but this would have resulted in considerably denser networks.

The revert networks were then pruned by removing self-reverts and by removing anonymous users, vandals, and bots, as well as any resulting isolates. We identified anonymous users as editors who did not use a registered username to revert (the data include an IP address instead) and vandals as editors who had all their edits reverted by others. The latter rule meant that we also got rid of newcomers who became discouraged and left Wikipedia after all their initial contributions were reverted. Since we are interested in social interactions emerging from repeated activity, we do not believe that this decision affects our results. To remove the bots, we identified all accounts with bot status (https://en.wikipedia.org/wiki/Wikipedia:Bots/Status) in the Wikipedia database (as of August 6, 2015). In addition, we removed any accounts with a username containing different spelling variations of the word "bot."

**Null model.** A number of sophisticated statistical methods for analyzing network data already exist but they are not adapted for large growing networks observed over a long continuous period of time. Exponential random-graph models (ERGMs) cannot account for the timing of interactions because they express the structural properties of aggregate networks or networks observed at a single moment in time (*42*). Extensions such as temporal ERGMs (*57*) and stochastic actor-based models (*58*) account for network dynamics but are nevertheless restricted to relatively small networks with a fixed set of nodes and a few "snapshots" over time.

To establish the statistical significance of our observations, we use a null model created by randomizing the underlying network. The randomization needs to preserve the daily pattern and the community structure of the network, while removing any systematic clustering in the timing of events in which an individual is involved, as we consider such clustering evidence for a social process. Hence, we do not randomize the network structure but only the timing of events. There are several possible ways to do this (*18*, *59*). First, one can shuffle the entire time sequence of events but this destroys individual activity patterns and increases the time variation in the individual's activities, as many editors joined Wikipedia for a limited period only and the shuffle is over almost 11 years. Individual activity patterns can be preserved if the randomization instead repeatedly samples two random nodes with equal number of events and swaps the time sequences for the events. The same could be achieved by shuffling within shorter periods of time, say 24 or 168 hours.

In addition to swapping and shuffling within a node, one can swap and shuffle within dyad, only within a node's in-links, or only within a node's out-links. These methods preserve the activity patterns characteristic for dyadic exchange, individual "visibility", and individual activity, respectively. However, in this way they also restrict the baseline to these narrow scopes. For example, swapping within dyads would compare the occurrence of the *AB-AB* motif to the *AB-BA* motif but would not answer whether either of them is more common than chance.

To account for these potential caveats, we choose to use as a baseline a shuffle of the timestamps of events within individuals and within a window of 24 hours. For each link *l*, we look at the source *i* and collect other links in which *i* participated (as either source or target) up to 24 hours before or after *l*. We then swap *l*'s time with the time of one of these links selected at random; if the set of candidate links is empty, no swap occurs. We shuffle within individuals to preserve individual activity patterns and to identify each social interaction against this expected sequence of events. We execute the shuffle within a limited time window to preserve the duration of activity per individuals. The time window should be at least 24 hours to allow for interactions between editors from different time zones and with different daily routines. We found that a time window on the order of 24-240 hours produces similar results. Hence, we chose 24 hours as this is the time window we use to define the social interactions.

In short, our shuffling method does not change the structure of the network (who reverts whom), which implies that it preserves the community structure centered around articles and topics. Neither does the method change the overall sequence of timestamps, thus preserving any natural burstiness of activity due to editors being in the same time zone or due to the occurrence of news-worthy events, for example. In contrast, the method deliberately shuffles the sequence of reverts that an editor is involved in, thus removing any individual behavioral patterns.

**Statistical tests.** The shuffled network provides us with an expected distribution for each interaction over the period of 24 hours. Comparing the weight of the distribution in the observed data with that in multiple shuffled realizations of the network can tell us the extent to which the interaction is more common than expected. Similarly, comparing the shape of the distribution can tell us the extent to which the interaction occurs at a faster rate than expected. To quantify the comparison, we estimate the *z*-score as follows:

$$Z = (X_D - \mu(X_R)) / \sigma(X_R). \tag{1}$$

In equation (1), $X_D$ is the relevant statistic in the data (for example, count or mean), $X_R$ is the same statistic in the shuffled networks, $\mu$ is the mean, and $\sigma$ is the standard deviation.

The statistic we use for frequency is the count of events (Supplementary Table S1). Two statistics we use for rate include the mean and skewness. Alternatively, to compare the shapes of two distributions, we also conduct a two-sample Kolmogorov-Smirnoff (KS) test between the data and one randomly chosen baseline generation. The KS test measures the maximum area deviation between the two normalized cumulative distributions. The sign of the KS statistic tells us the direction of the difference while the

*p*-value tells us the likelihood of the observed distance given the null hypothesis of the distributions being the same. The problem with the KS test is that it is extremely sensitive to small deviations and thus overestimates the significance of the differences. Hence, we use the signed KS statistic to report the results but we use the mean and skewness *z*-scores to discuss their significance. Overall, the results are similar in terms of direction across the three statistics (Supplementary Table S2).

**Status.** To investigate the effect of status in the interactions, we operationalize status as the base-ten logarithm of the number of edits the editor has completed by the time of the revert under question. We transformed the number of edits with the logarithm because they follow a power-law distribution (Supplementary Fig. S18). As a result, the difference in status can then also be expressed as the base-ten logarithm of the ratio of number of edits. To find evidence that status plays a role, we need to compare the status difference in the focal interaction with the status difference in any other revert. The observations are not independent across and within these two groups, however, but nested within individuals. To account for this, we execute regression analyses with standard errors adjusted for clustering within reverter and reverted (Supplementary Table S3).

**Acknowledgements**

This project has received funding from the European Union's Horizon 2020 research and innovation program under grant agreement No 645043. The authors thank Wikimedia Deutchland e.V. and Wikimedia Foundation for the live access to the Wikipedia data via Toolserver.

**Author contributions**

M.T. and T.Y. designed and performed research. M.T. analyzed data. M.T., R.G., and T.Y. wrote and reviewed the manuscript.

**Supporting Information for:**

**Dynamics of Disagreement: Large-Scale Temporal Network Analysis Reveals Negative Interactions in Online Collaboration**

Milena Tsvetkova, Ruth García-Gavilanes, and Taha Yasseri

**Table S1. Results for the counts of the six motifs for 13 different language editions of Wikipedia.**

| Language | AB-AB Count data/null | Z | AB-BA Count data/null | Z | AB-BC Count data/null | Z | AB-CB Count data/null | Z | AB-AC Count data/null | Z | AB-CA Count data/null | Z |
|---|---|---|---|---|---|---|---|---|---|---|---|---|
| English | 1018464/1004006 | 74.4* | 539659/523510 | 49.6* | 611112/717717 | -204.9* | 966610/988516 | -36.3* | 1723351/1779957 | -89.0* | 1242189/1185190 | 76.2* |
| Spanish | 83589/81208 | 33.8* | 36548/34373 | 20.9* | 38351/46190 | -39.3* | 57297/58629 | -8.1* | 131561/136766 | -21.9* | 86382/81164 | 15.2* |
| German | 28926/28654 | 13.2* | 24463/22587 | 22.9* | 10977/16383 | -19.5* | 21061/21390 | -4.4* | 24235/25819 | -14.2* | 21939/18095 | 13.5* |
| Japanese | 59833/59539 | 10.7* | 21562/23032 | -13.2* | 11333/19166 | -49.2* | 25581/26784 | -12.5* | 32332/35026 | -16.8* | 24039/21739 | 11.0* |
| French | 35759/35400 | 11.9* | 17180/16297 | 12.7* | 9076/12904 | -22.6* | 18949/19375 | -5.3* | 26677/28883 | -15.6* | 18142/15755 | 11.8* |
| Portuguese | 56504/54762 | 26.1* | 28549/27038 | 14.9* | 24935/29715 | -29.8* | 34839/35727 | -5.0* | 82530/86292 | -21.2* | 53664/51893 | 7.2* |
| Chinese | 46274/45704 | 14.7* | 22779/23113 | -5.7* | 15918/21862 | -24.3* | 23773/25127 | -8.9* | 44061/47358 | -17.3* | 30481/29103 | 7.5* |
| Hebrew | 22460/21770 | 23.3* | 11645/10530 | 24.8* | 13423/16279 | -25.7* | 19189/19691 | -4.4* | 33904/35734 | -15.0* | 23947/22528 | 11.8* |
| Arabic | 7857/7750 | 8.7* | 3676/3511 | 6.4* | 2656/3483 | -24.4* | 4276/4410 | -2.2 | 11901/12538 | -15.4* | 5288/4893 | 5.1* |
| Hungarian | 6144/6065 | 6.6* | 2535/2304 | 6.1* | 1853/2523 | -23.2* | 3508/3640 | -2.4 | 6734/7116 | -9.6* | 3636/3235 | 9.7* |
| Persian | 11642/11389 | 13.0* | 7859/7236 | 13.0* | 7358/9079 | -22.8* | 10045/10152 | -1.4 | 17512/18319 | -8.8* | 12522/11722 | 6.2* |
| Czech | 5097/5035 | 6.5* | 2524/2308 | 10.2* | 1806/2400 | -17.9* | 3297/3236 | 1.3 | 4975/5144 | -4.8* | 3129/2842 | 6.6* |
| Romanian | 4197/4106 | 10.2* | 1412/1418 | -0.2 | 999/1369 | -10.8* | 1861/1963 | -2.4 | 5759/5889 | -2.9* | 2758/2527 | 4.0* |

* 2-sided p-value < 0.01

**Table S2. Results for the response rate of the six motifs for 13 different language editions of Wikipedia.**

| Language | AB-AB | | | | | AB-BA | | | | | AB-BC | | | | |
|---|---|---|---|---|---|---|---|---|---|---|---|---|---|---|---|
| | KS | Mean data /null | Mean Z | Skew data /null | Skew Z | KS | Mean data /null | Mean Z | Skew data /null | Skew Z | KS | Mean data /null | Mean Z | Skew data /null | Skew Z |
| English | 0.20* | 100/221 | -459.5* | 3.26/1.79 | 963.8* | 0.08* | 241/313 | -99.0* | 1.69/1.27 | 98.1* | -0.13* | 514/404 | 142.6* | 0.52/0.88 | -97.4* |
| Spanish | 0.25* | 81/226 | -184.3* | 3.72/1.75 | 343.9* | 0.09* | 242/327 | -49.4* | 1.66/1.2 | 55.3* | -0.16* | 511/381 | 64.3* | 0.53/0.97 | -47.2* |
| German | 0.10* | 128/190 | -47.0* | 2.81/2.06 | 59.4* | 0.12* | 195/285 | -37.9* | 2.04/1.44 | 41.4* | -0.19* | 544/381 | 50.8* | 0.45/1.00 | -36.6* |
| Japanese | 0.13* | 56/122 | -112.0* | 4.62/2.81 | 177.6* | -0.16* | 274/239 | 13.4* | 1.44/1.67 | -11.8* | -0.30* | 513/295 | 62.3* | 0.56/1.34 | -41.7* |
| French | 0.12* | 85/159 | -69.2* | 3.65/2.37 | 98.6* | 0.07* | 211/275 | -19.5* | 1.95/1.5 | 24.6* | -0.19* | 560/390 | 40.3* | 0.38/0.95 | -28.6* |
| Portuguese | 0.26* | 89/238 | -112.3* | 3.53/1.68 | 195.1* | 0.10* | 210/306 | -32.1* | 1.89/1.3 | 36.3* | -0.19* | 510/363 | 60.7* | 0.56/1.05 | -43.5* |
| Chinese | 0.19* | 91/184 | -83.2* | 3.47/2.08 | 133.1* | -0.15* | 300/273 | 8.7* | 1.32/1.46 | -8.0* | -0.26* | 557/336 | 45.5* | 0.35/1.15 | -34.7* |
| Hebrew | 0.24* | 76/230 | -78.3* | 3.88/1.70 | 145.5* | 0.25* | 151/336 | -29.7* | 2.47/1.15 | 47.7* | -0.16* | 516/403 | 28.4* | 0.52/0.88 | -19.8* |
| Arabic | 0.18* | 125/233 | -36.7* | 2.80/1.69 | 47.4* | -0.07* | 319/354 | -5.2* | 1.22/1.07 | 4.5* | -0.22* | 623/431 | 22.5* | 0.17/0.77 | -15.5* |
| Hungarian | 0.15* | 75/173 | -39.3* | 3.90/2.20 | 61.6* | 0.08* | 229/312 | -11.8* | 1.80/1.29 | 13.3* | -0.22* | 594/417 | 16.1* | 0.29/0.84 | -11.9* |
| Persian | 0.24* | 126/267 | -64.3* | 2.82/1.49 | 86.1* | 0.19* | 222/361 | -30.9* | 1.81/1.04 | 40.3* | -0.15* | 532/399 | 22.3* | 0.47/0.90 | -16.0* |
| Czech | 0.16* | 76/173 | -33.0* | 3.90/2.20 | 55.7* | 0.07* | 234/280 | -4.4* | 1.79/1.46 | 5.6* | -0.22* | 566/401 | 15.9* | 0.38/0.89 | -10.3* |
| Romanian | 0.19* | 59/181 | -37.6* | 4.61/2.16 | 73.4* | -0.09* | 211/264 | -4.7* | 1.90/1.57 | 4.2* | -0.24* | 487/324 | 14.0* | 0.63/1.24 | -10.0* |

* 2-sided p-value < 0.01

| Language | AB-CB | | | | | AB-AC | | | | | AB-CA | | | | |
|---|---|---|---|---|---|---|---|---|---|---|---|---|---|---|---|
| | KS | Mean data /null | Mean Z | Skew data /null | Skew Z | KS | Mean data /null | Mean Z | Skew data /null | Skew Z | KS | Mean data /null | Mean Z | Skew data /null | Skew Z |
| English | 0.09* | 374/439 | -126.9* | 1.03/0.78 | 230.8* | -0.12* | 380/323 | 146.0* | 1.00/1.23 | -132.9* | -0.05* | 410/428 | -15.1* | 0.94/0.83 | 26.4* |
| Spanish | 0.07* | 406/464 | -29.7* | 0.89/0.68 | 27.6* | -0.15* | 379/303 | 40.3* | 1.00/1.32 | -34.1* | -0.10* | 459/443 | 4.5* | 0.75/0.75 | -0.4 |
| German | 0.02* | 364/381 | -6.9* | 1.10/1.01 | 8.0* | -0.11* | 492/402 | 36.7* | 0.60/0.91 | -28.4* | 0.04* | 357/393 | -8.8* | 1.13/0.95 | 10.9* |
| Japanese | -0.02* | 338/345 | -2.2 | 1.16/1.12 | 2.1 | -0.21* | 413/293 | 37.0* | 0.86/1.36 | -28.3* | -0.17* | 413/350 | 10.0* | 0.87/1.08 | -7.6* |
| French | 0.05* | 355/388 | -8.4* | 1.11/0.97 | 8.2* | -0.18* | 487/364 | 44.0* | 0.62/1.05 | -34.3* | -0.10* | 418/406 | 1.8 | 0.88/0.89 | -0.2 |
| Portuguese | 0.04* | 442/478 | -10.6* | 0.77/0.64 | 10.2* | -0.16* | 377/293 | 58.8* | 1.05/1.39 | -39.9* | -0.09* | 483/454 | 8.2* | 0.67/0.72 | -3.5* |
| Chinese | -0.05* | 437/418 | 2.5 | 0.75/0.8 | -1.3 | -0.14* | 396/300 | 39.8* | 0.92/1.32 | -31.1 | -0.16* | 446/405 | 8.2* | 0.78/0.87 | -4.7* |
| Hebrew | 0.10* | 393/461 | -14.6* | 0.95/0.69 | 14.6* | -0.15* | 441/361 | 29.8* | 0.77/1.04 | -22.9* | -0.04* | 443/453 | -2.6* | 0.77/0.70 | 4.2* |
| Arabic | -0.05* | 509/504 | 0.8 | 0.58/0.57 | 0.5 | -0.09* | 400/331 | 17.3* | 0.89/1.16 | -15.0* | -0.11* | 515/493 | 2.5 | 0.58/0.56 | 0.7 |
| Hungarian | 0.05* | 404/430 | -4.1* | 0.89/0.79 | 3.1* | -0.13* | 431/353 | 23.9* | 0.79/1.08 | -20.1* | -0.05* | 438/450 | -1.0 | 0.81/0.72 | 1.9 |
| Persian | 0.05* | 402/446 | -9.7* | 0.93/0.74 | 9.3* | -0.13* | 404/338 | 18.2* | 0.91/1.15 | -13.6* | -0.04* | 419/441 | -2.9* | 0.86/0.75 | 4.3* |
| Czech | 0.07* | 376/420 | -3.4* | 0.99/0.82 | 4.0* | -0.19* | 509/393 | 15.7* | 0.55/0.93 | -10.8* | -0.05* | 444/451 | -0.4 | 0.79/0.72 | 1.3 |
| Romanian | 0.09* | 371/428 | -4.2* | 1.07/0.79 | 3.9* | -0.17* | 360/282 | 17.6* | 1.13/1.46 | -11.8* | -0.09* | 460/451 | 0.6 | 0.78/0.76 | 0.2 |

* 2-sided p-value < 0.01

**Table S3. Results for the status difference in the six motifs for 13 different language editions of Wikipedia.**

| Language | AB-AB | | AB-BA | | AB-BC | | | AB-CB | | | AB-AC | | | AB-CA | | |
|---|---|---|---|---|---|---|---|---|---|---|---|---|---|---|---|---|
| | Other | A-B | Other | A-B | Other | A-B | B-C | Other | A-B | C-B | Other | A-B | A-C | Other | A-B | C-A |
| English | 1.069* | 0.121* | 1.396* | -1.108* | 1.428* | -1.723* | -0.173* | 1.293* | -0.485* | -0.282* | 0.535* | 1.078* | 0.648* | 1.277* | 0.075 | -1.460* |
| | (0.023) | (0.024) | (0.026) | (0.023) | (0.025) | (0.028) | (0.017) | (0.025) | (0.024) | (0.021) | (0.017) | (0.030) | (0.010) | (0.019) | (0.033) | (0.020) |
| Spanish | 1.223* | 0.338* | 1.629* | -1.119* | 1.674* | -2.134* | -0.086 | 1.554* | -0.647* | -0.495* | 0.675* | 1.240* | 0.839* | 1.493* | 0.220 | -1.818* |
| | (0.082) | (0.079) | (0.077) | (0.103) | (0.075) | (0.126) | (0.058) | (0.081) | (0.133) | (0.054) | (0.077) | (0.101) | (0.038) | (0.051) | (0.152) | (0.084) |
| German | 0.789* | 0.025 | 1.022* | -0.719* | 0.881* | -0.933* | -0.363* | 0.795* | 0.019 | 0.007 | 0.720* | 0.331* | 0.301* | 0.946* | -0.602* | -0.755* |
| | (0.039) | (0.046) | (0.041) | (0.044) | (0.041) | (0.060) | (0.049) | (0.042) | (0.047) | (0.031) | (0.036) | (0.066) | (0.040) | (0.039) | (0.051) | (0.044) |
| Japanese | 0.636* | 0.088 | 0.855* | -0.769* | 0.773* | -1.034* | -0.135 | 0.661* | 0.158* | 0.098 | 0.587* | 0.367* | 0.380* | 0.813* | -0.520* | -0.671* |
| | (0.059) | (0.061) | (0.070) | (0.071) | (0.067) | (0.107) | (0.078) | (0.069) | (0.061) | (0.061) | (0.055) | (0.091) | (0.056) | (0.061) | (0.098) | (0.090) |
| French | 1.026* | 0.192* | 1.348* | -1.045* | 1.219* | -1.282* | -0.339* | 1.097* | 0.091 | 0.009 | 0.953* | 0.547* | 0.445* | 1.271* | -0.555* | -1.123* |
| | (0.057) | (0.042) | (0.058) | (0.057) | (0.059) | (0.076) | (0.060) | (0.063) | (0.039) | (0.036) | (0.050) | (0.091) | (0.043) | (0.056) | (0.074) | (0.064) |
| Portuguese | 1.338* | 0.013 | 1.699* | -1.301* | 1.695* | -2.201* | -0.057 | 1.552* | -0.690* | -0.507* | 0.580* | 1.344* | 0.960* | 1.504* | 0.148 | -1.870* |
| | (0.125) | (0.090) | (0.098) | (0.164) | (0.092) | (0.162) | (0.075) | (0.107) | (0.174) | (0.089) | (0.127) | (0.104) | (0.062) | (0.087) | (0.109) | (0.127) |
| Chinese | 0.992* | -0.146 | 1.200* | -0.966* | 1.134* | -1.508* | -0.016 | 1.049* | -0.441* | -0.438* | 0.503* | 0.966* | 0.661* | 1.053* | -0.069 | -1.241* |
| | (0.226) | (0.122) | (0.186) | (0.100) | (0.168) | (0.192) | (0.120) | (0.192) | (0.158) | (0.133) | (0.101) | (0.258) | (0.112) | (0.119) | (0.256) | (0.173) |
| Hebrew | 1.081* | 0.177 | 1.416* | -0.987* | 1.438* | -1.608* | -0.240 | 1.255* | -0.281 | -0.259* | 0.700* | 0.909* | 0.639* | 1.373* | -0.145 | -1.380* |
| | (0.132) | (0.078) | (0.119) | (0.111) | (0.113) | (0.143) | (0.094) | (0.125) | (0.140) | (0.084) | (0.120) | (0.100) | (0.057) | (0.110) | (0.111) | (0.117) |
| Arabic | 1.450* | -0.167 | 1.648* | -1.401* | 1.562* | -1.718* | -0.384* | 1.446* | -0.257 | -0.337* | 0.925* | 0.899* | 0.731* | 1.583* | -0.362* | -1.665* |
| | (0.163) | (0.077) | (0.139) | (0.148) | (0.144) | (0.143) | (0.083) | (0.164) | (0.105) | (0.086) | (0.112) | (0.149) | (0.083) | (0.134) | (0.107) | (0.141) |
| Hungarian | 1.241* | -0.018 | 1.440* | -1.007* | 1.379* | -1.499* | -0.465* | 1.260* | -0.068 | -0.109 | 0.969* | 0.775* | 0.516* | 1.399* | -0.403 | -1.304* |
| | (0.157) | (0.084) | (0.144) | (0.131) | (0.143) | (0.136) | (0.087) | (0.156) | (0.134) | (0.111) | (0.110) | (0.212) | (0.085) | (0.131) | (0.165) | (0.113) |
| Persian | 1.096* | 0.040 | 1.525* | -1.190* | 1.478* | -1.617* | -0.528* | 1.304* | -0.507* | -0.537* | 0.644* | 0.782* | 0.544* | 1.513* | -0.580* | -1.573* |
| | (0.163) | (0.092) | (0.135) | (0.107) | (0.140) | (0.126) | (0.077) | (0.164) | (0.118) | (0.075) | (0.159) | (0.105) | (0.069) | (0.133) | (0.109) | (0.119) |
| Czech | 1.004* | 0.292 | 1.319* | -0.881* | 1.239* | -1.289* | -0.246 | 1.097* | 0.083 | -0.022 | 0.849* | 0.797* | 0.546* | 1.279* | -0.481 | -1.115* |
| | (0.163) | (0.113) | (0.164) | (0.147) | (0.169) | (0.202) | (0.121) | (0.187) | (0.100) | (0.088) | (0.122) | (0.208) | (0.123) | (0.151) | (0.197) | (0.151) |
| Romanian | 1.614* | -0.126 | 1.800* | -1.091* | 1.761* | -2.429* | -0.057 | 1.631* | -0.240 | -0.323 | 0.958* | 1.195* | 0.957* | 1.713* | 0.011 | -1.965* |
| | (0.335) | (0.195) | (0.242) | (0.195) | (0.237) | (0.380) | (0.185) | (0.275) | (0.211) | (0.283) | (0.217) | (0.226) | (0.203) | (0.222) | (0.271) | (0.415) |

\* 2-sided p-value < 0.01
The table reports coefficients from ordinary least-square regression predicting the difference in the log with base ten of the number of edits completed by the time of the revert by the reverter and the reverted individual. The standards errors (in brackets) account for two-way clustering within reverter and reverted.

**Table S4. Results for the proportion of the six motifs that occurred in the same article for 13 different language editions of Wikipedia.**

| | AB-AB | | AB-BA | | AB-BC | | AB-CB | | AB-AC | | AB-CA | |
|---|---|---|---|---|---|---|---|---|---|---|---|---|
| Language | Prop. data /rand | Z | Prop. data /rand | Z | Prop. data /rand | Z | Prop. data /rand | Z | Prop. data /rand | Z | Prop. data /rand | Z |
| English | 0.222 /0.180 | 137.3* | 0.685 /0.180 | 1043.7* | 0.098 /0.180 | -188.7* | 0.171 /0.180 | -26.6* | 0.055 /0.180 | -690.0* | 0.154 /0.180 | -95.5* |
| Spanish | 0.204 /0.150 | 53.7* | 0.677 /0.150 | 317.5* | 0.075 /0.150 | -46.1* | 0.155 /0.149 | 4.8* | 0.029 /0.150 | -199.7* | 0.107 /0.150 | -45.2* |
| German | 0.411 /0.451 | -19.1* | 0.884 /0.451 | 179.3* | 0.289 /0.451 | -38.9* | 0.359 /0.451 | -30.2* | 0.185 /0.450 | -101.1* | 0.476 /0.451 | 8.3* |
| Japanese | 0.094 /0.181 | -86.5* | 0.572 /0.182 | 173.6* | 0.098 /0.182 | -23.1* | 0.124 /0.181 | -26.1* | 0.054 /0.181 | -73.5* | 0.212 /0.182 | 13.1* |
| French | 0.231 /0.274 | -27.5* | 0.762 /0.273 | 160.0* | 0.179 /0.273 | -22.8* | 0.240 /0.274 | -12.6* | 0.082 /0.274 | -76.8* | 0.291 /0.274 | 5.3* |
| Portuguese | 0.227 /0.159 | 54.5* | 0.652 /0.158 | 251.8* | 0.069 /0.159 | -48.6* | 0.151 /0.159 | -4.2* | 0.029 /0.159 | -154.3* | 0.103 /0.159 | -44.2* |
| Chinese | 0.136 /0.167 | -28.7* | 0.528 /0.166 | 173.9* | 0.069 /0.167 | -37.5* | 0.112 /0.166 | -29.4* | 0.038 /0.166 | -95.4* | 0.128 /0.167 | -19.7* |
| Hebrew | 0.248 /0.195 | 25.1* | 0.760 /0.195 | 146.7* | 0.096 /0.194 | -32.5* | 0.192 /0.195 | -1.0 | 0.051 /0.195 | -95.6* | 0.193 /0.195 | -1.0 |
| Arabic | 0.188 /0.160 | 8.3* | 0.673 /0.160 | 86.6* | 0.091 /0.160 | -10.5* | 0.151 /0.160 | -1.8 | 0.033 /0.160 | -60.4* | 0.165 /0.160 | 1.3 |
| Hungarian | 0.182 /0.194 | -2.9* | 0.728 /0.193 | 70.9* | 0.143 /0.195 | -6.1* | 0.220 /0.195 | 4.5* | 0.056 /0.195 | -38.9* | 0.261 /0.196 | 11.0* |
| Persian | 0.271 /0.216 | 19.0* | 0.682 /0.216 | 111.0* | 0.126 /0.216 | -20.0* | 0.220 /0.216 | 1.6 | 0.071 /0.216 | -64.8* | 0.242 /0.216 | 7.9* |
| Czech | 0.185 /0.231 | -10.1* | 0.737 /0.231 | 68.7* | 0.116 /0.231 | -11.4* | 0.237 /0.231 | 1.0 | 0.054 /0.231 | -34.9* | 0.284 /0.231 | 8.0* |
| Romanian | 0.176 /0.143 | 7.2* | 0.657 /0.142 | 60.5* | 0.086 /0.142 | -5.6* | 0.212 /0.144 | 8.8* | 0.020 /0.143 | -40.4* | 0.147 /0.143 | 0.6 |

* 2-sided p-value < 0.01

The construct the baseline, we first matched every revert in the network *ij* with another revert randomly selected from the set of reverts that *i* participated in (either as a reverter or reverted) within a time window of 24 hours. We then repeatedly sampled from this set of matched pairs. The set of samples was used to estimate the Z-score for the observed proportion of same-article interactions.

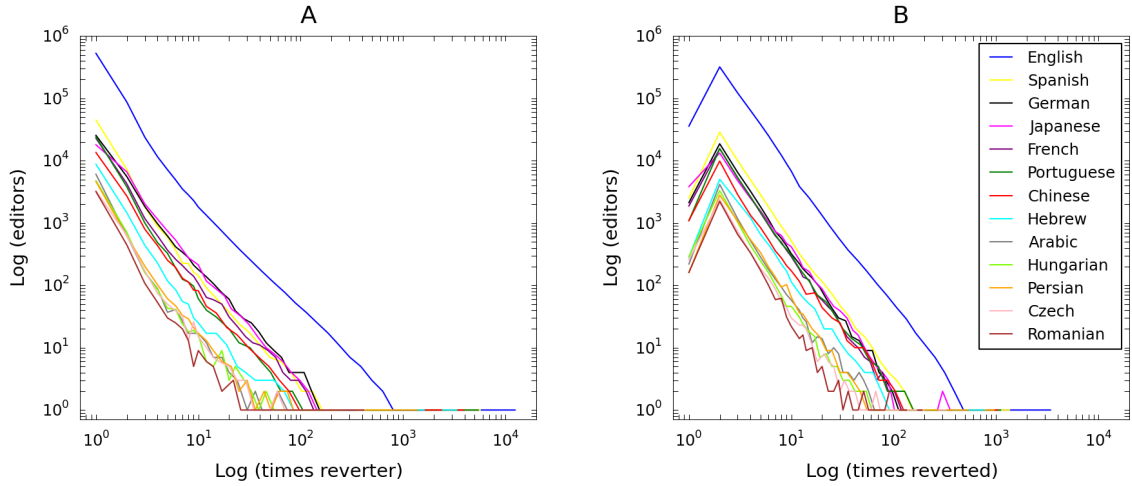

**Fig. S1.** Out-degree (A) and in-degree (B) distributions in the 13 networks of reverts.

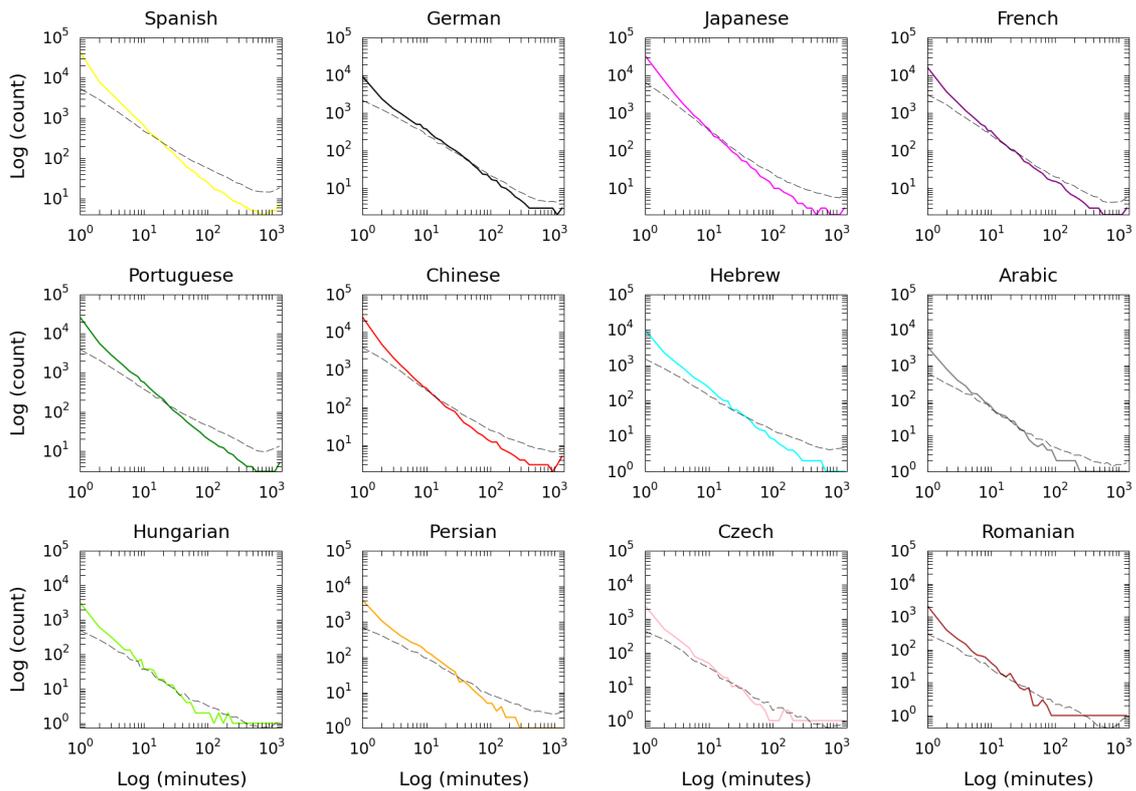

**Fig. S2.** Expected and observed counts of the *AB-AB* motif for 12 different language editions of Wikipedia. The dashed lines show the expected distributions according to the null model. Error bars are not shown as they are too small.

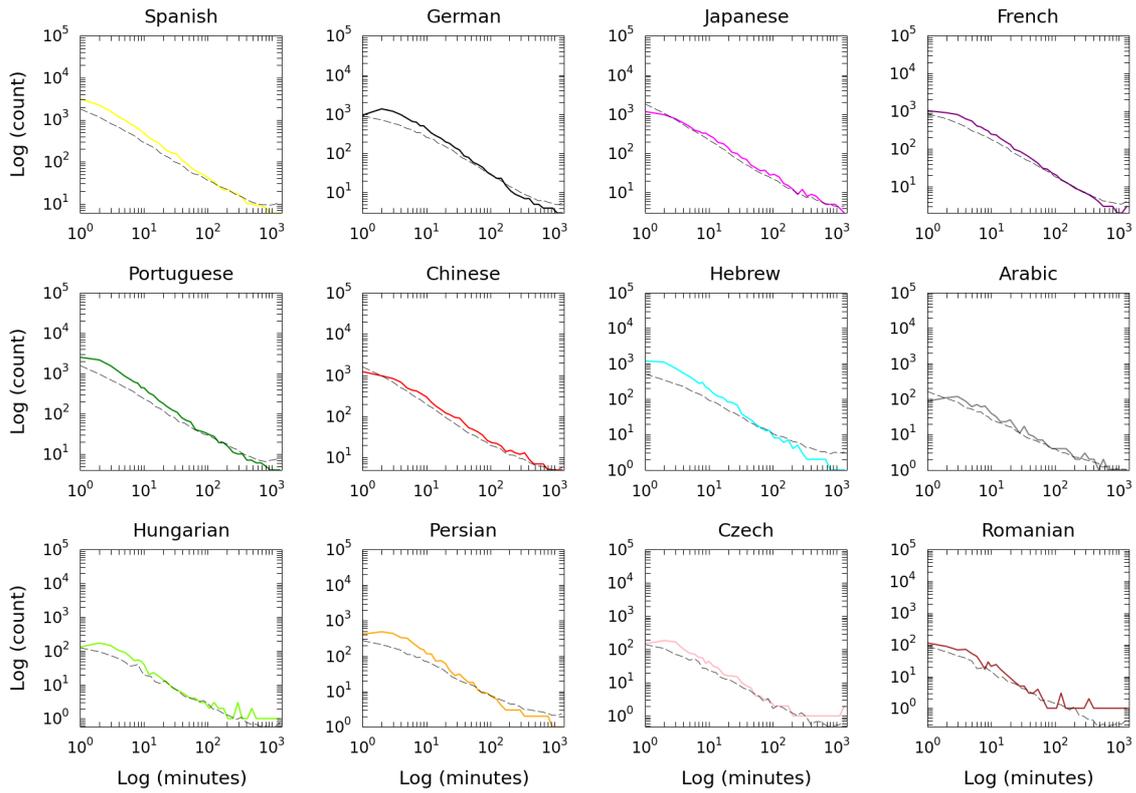

**Fig. S3.** Expected and observed counts of the *AB-BA* motif for 12 different language editions of Wikipedia. The dashed lines show the expected distributions according to the null model. Error bars are not shown as they are too small.

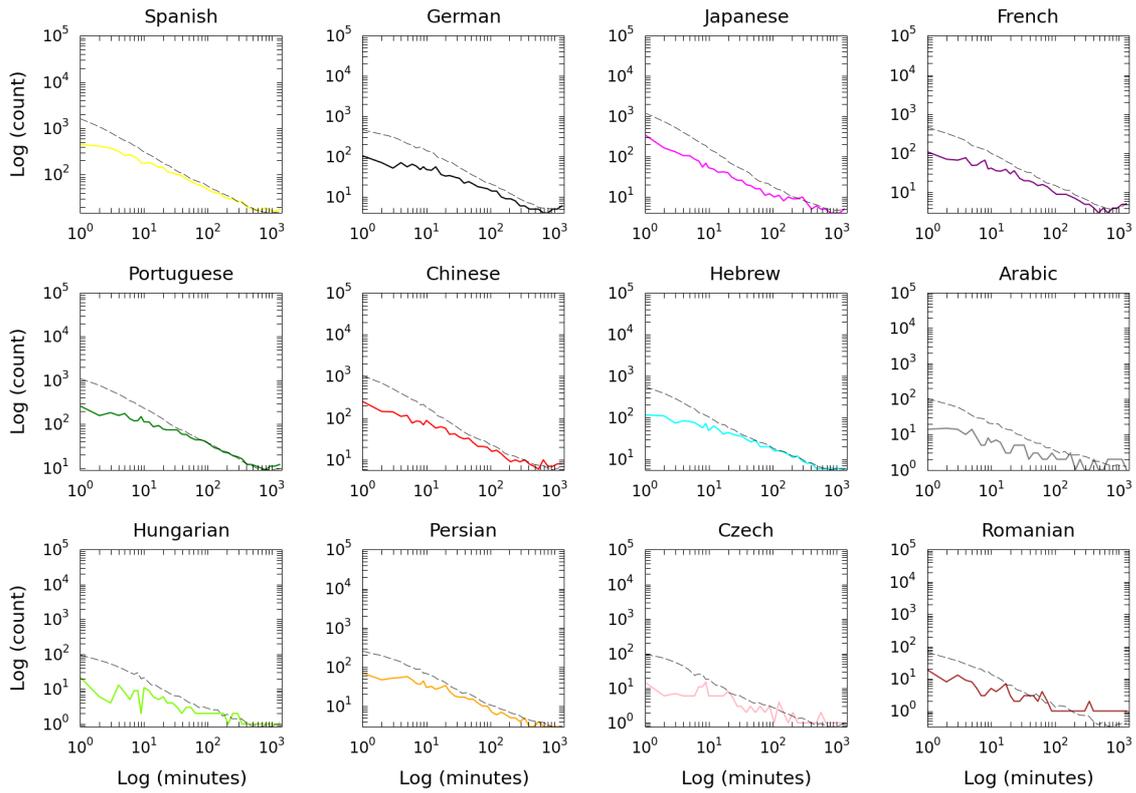

**Fig. S4.** Expected and observed counts of the *AB-BC* motif for 12 different language editions of Wikipedia. The dashed lines show the expected distributions according to the null model. Error bars are not shown as they are too small.

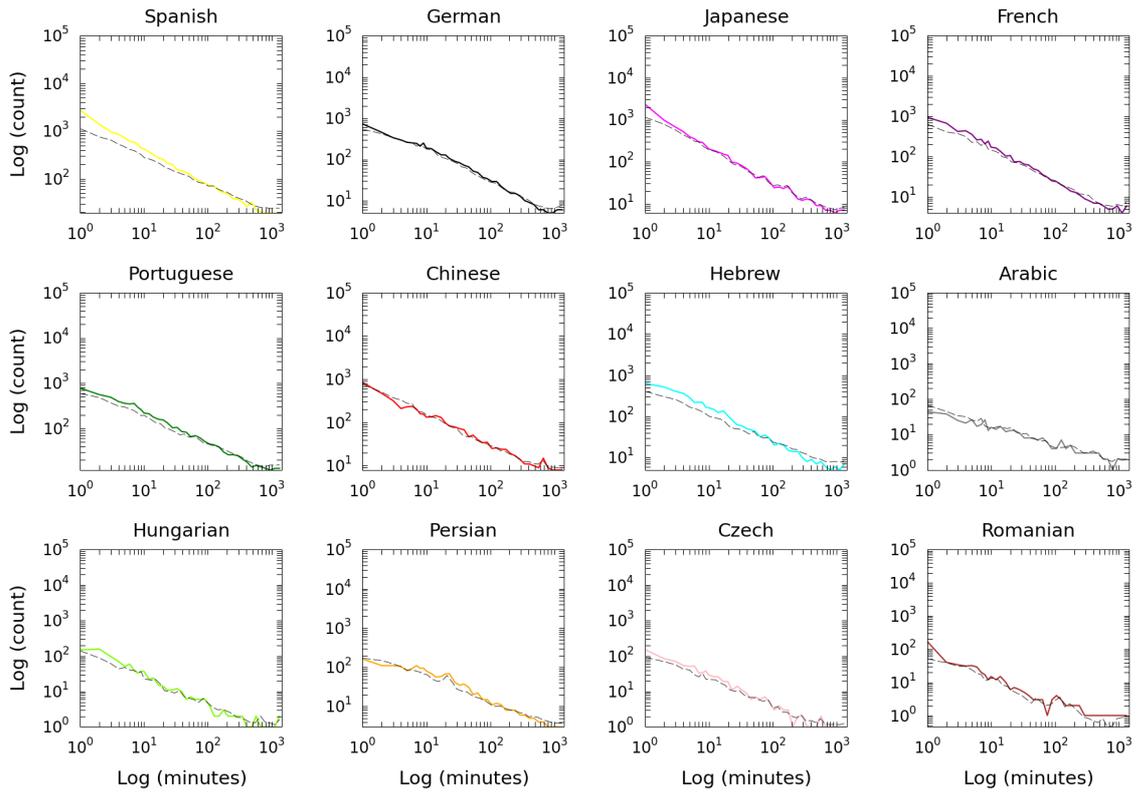

**Fig. S5.** Expected and observed counts of the *AB-CB* motif for 12 different language editions of Wikipedia. The dashed lines show the expected distributions according to the null model. Error bars are not shown as they are too small.

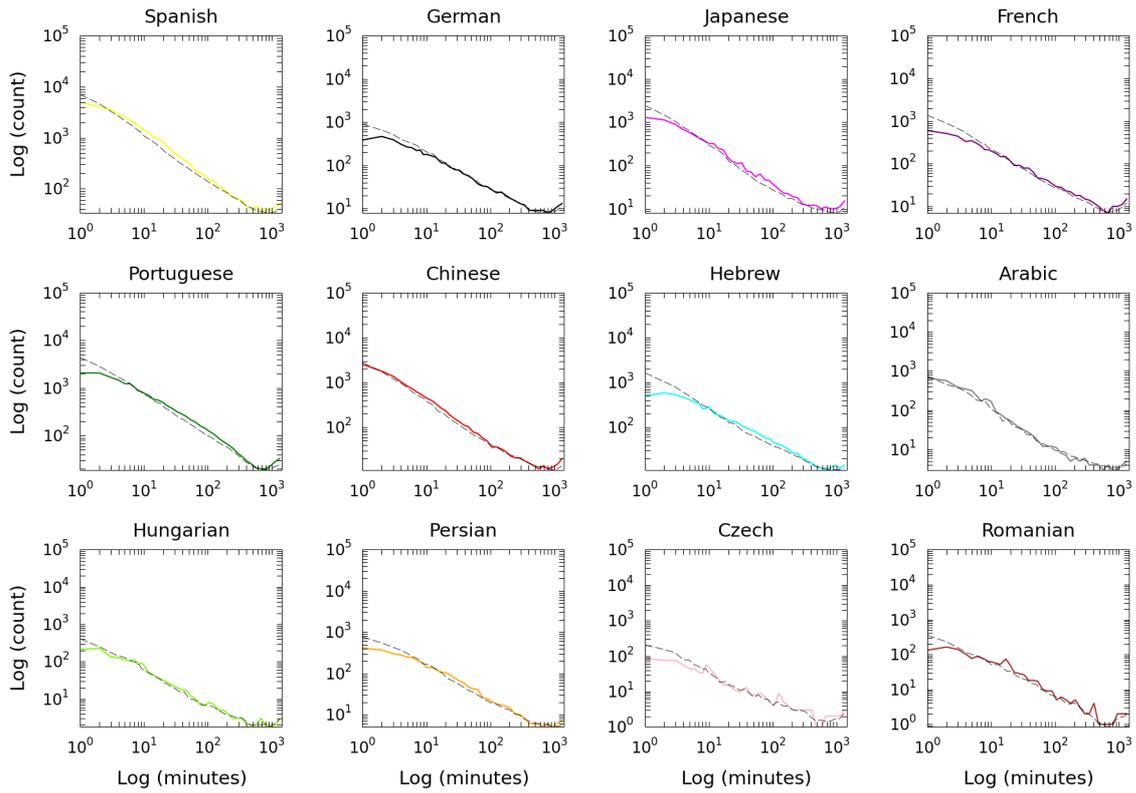

**Fig. S6.** Expected and observed counts of the *AB-AC* motif for 12 different language editions of Wikipedia. The dashed lines show the expected distributions according to the null model. Error bars are not shown as they are too small.

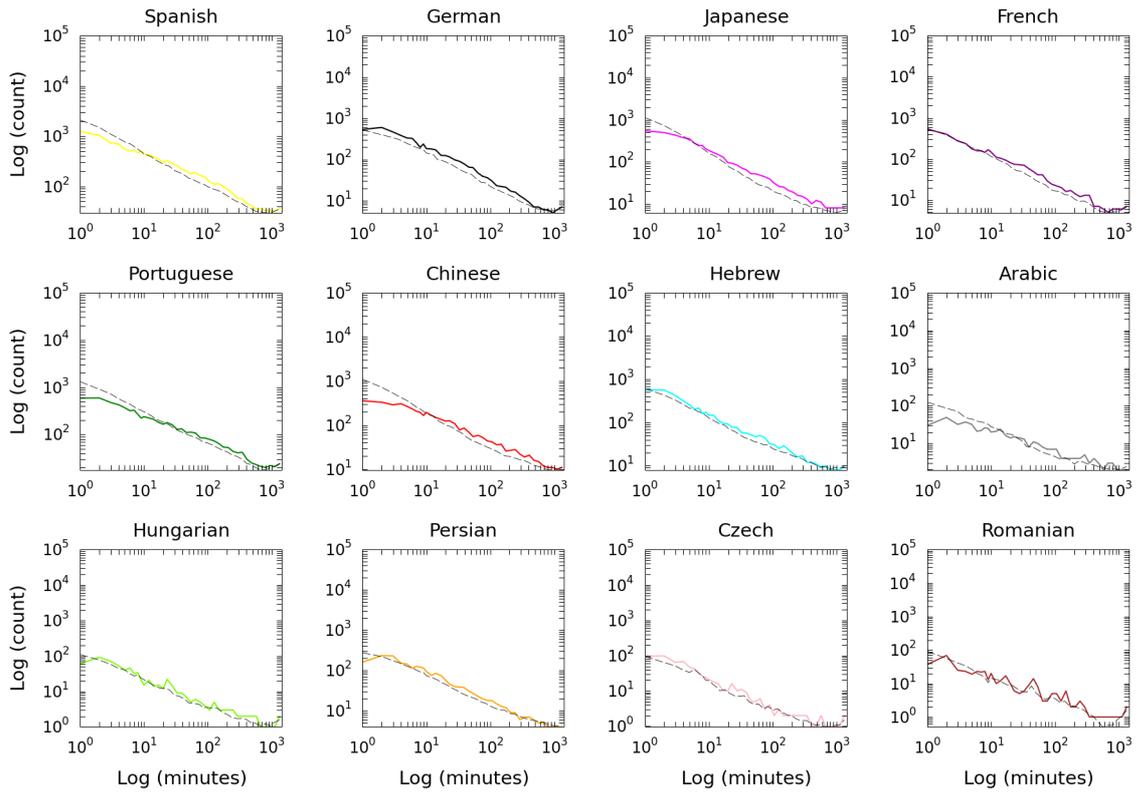

**Fig. S7.** Expected and observed counts of the *AB-CA* motif for 12 different language editions of Wikipedia. The dashed lines show the expected distributions according to the null model. Error bars are not shown as they are too small.

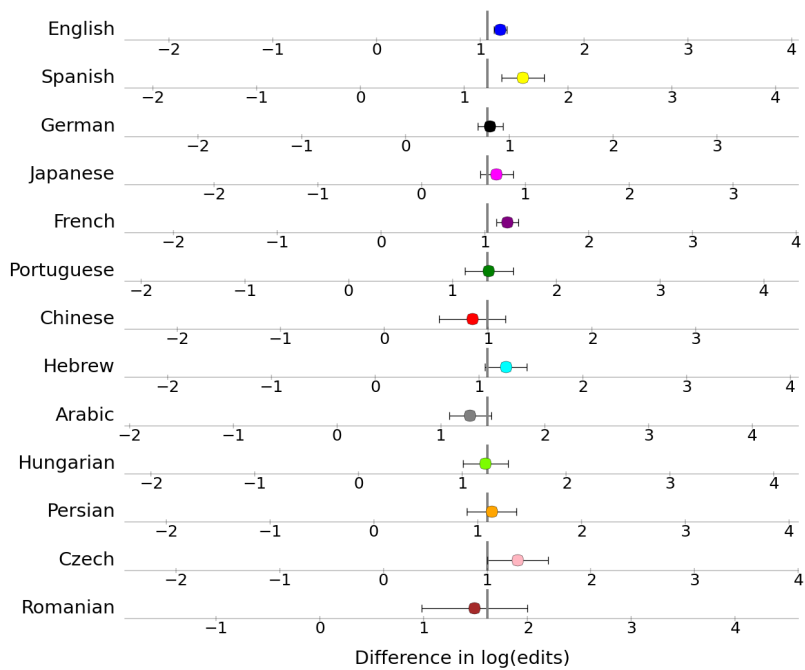

**Fig. S8.** Expected and observed difference in status between *A* and *B* for the *AB-AB* motif. Gray vertical lines show the expectation. Confidence intervals show 2.6 standard errors, corresponding to *p* < 0.01.

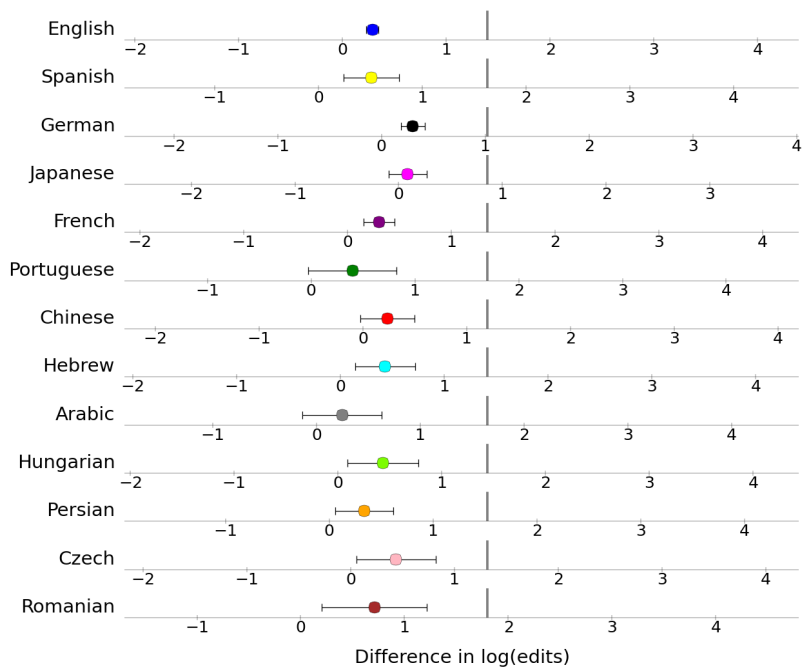

**Fig. S9.** Expected and observed difference in status between *A* and *B* for the *AB-BA* motif. Gray vertical lines show the expectation. Confidence intervals show 2.6 standard errors, corresponding to *p* < 0.01.

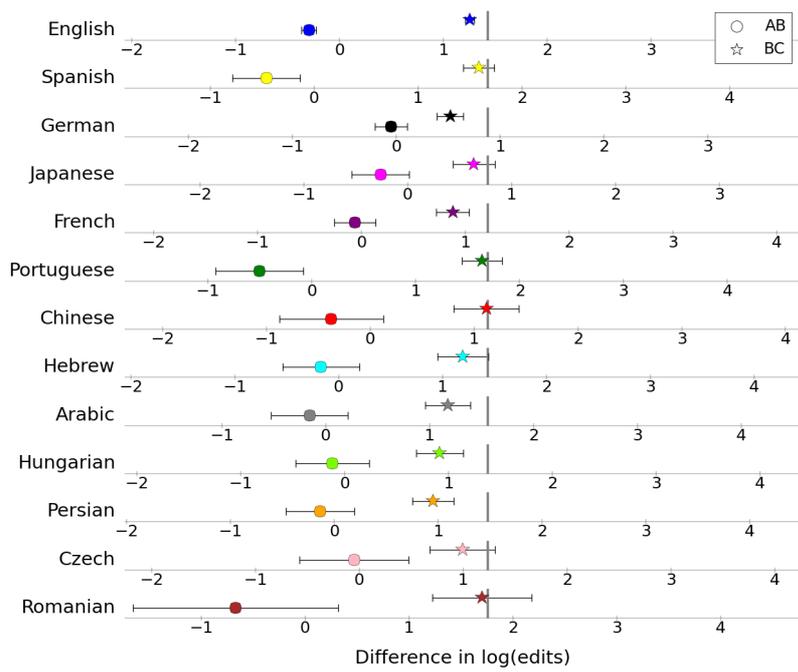

**Fig. S10.** Expected and observed difference in status between *A* and *B* (circles) and *B* and *C* (stars) for the *AB-BC* motif. Gray vertical lines show the expectation. Confidence intervals show 2.6 standard errors, corresponding to $p < 0.01$.

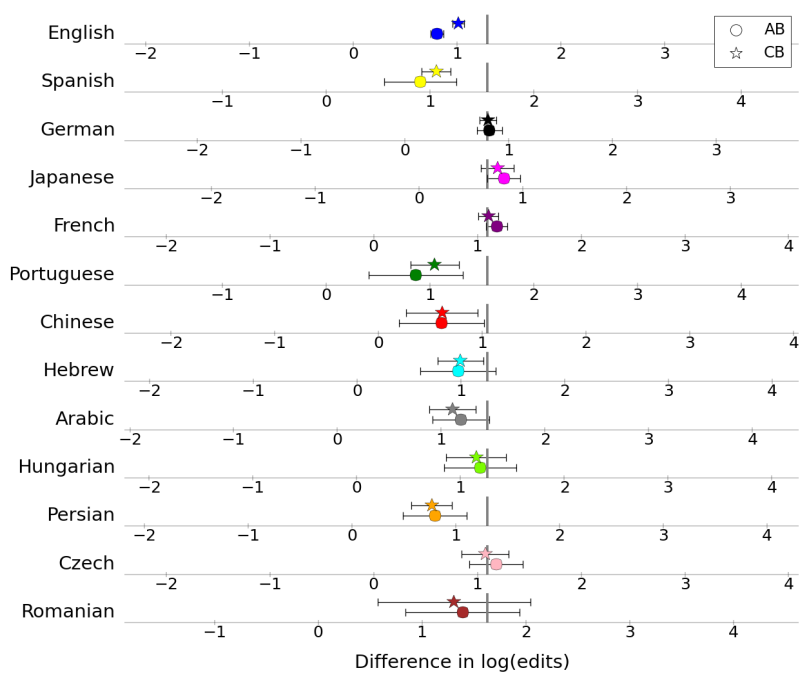

**Fig. S11.** Expected and observed difference in status between *A* and *B* (circles) and *C* and *B* (stars) for the *AB-CB* motif. Gray vertical lines show the expectation. Confidence intervals show 2.6 standard errors, corresponding to $p < 0.01$.

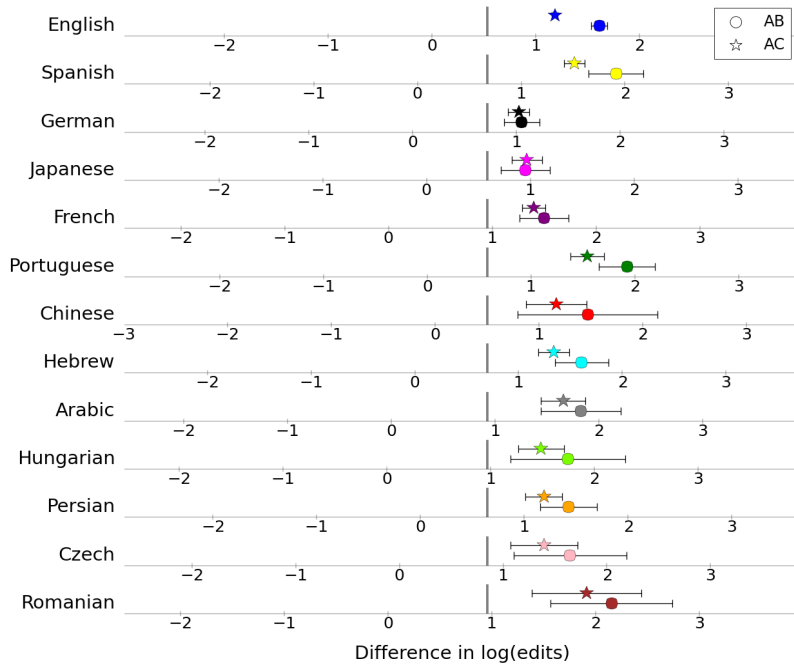

**Fig. S12.** Expected and observed difference in status between *A* and *B* (circles) and *A* and *C* (stars) for the *AB-AC* motif. Gray vertical lines show the expectation. Confidence intervals show 2.6 standard errors, corresponding to $p < 0.01$.

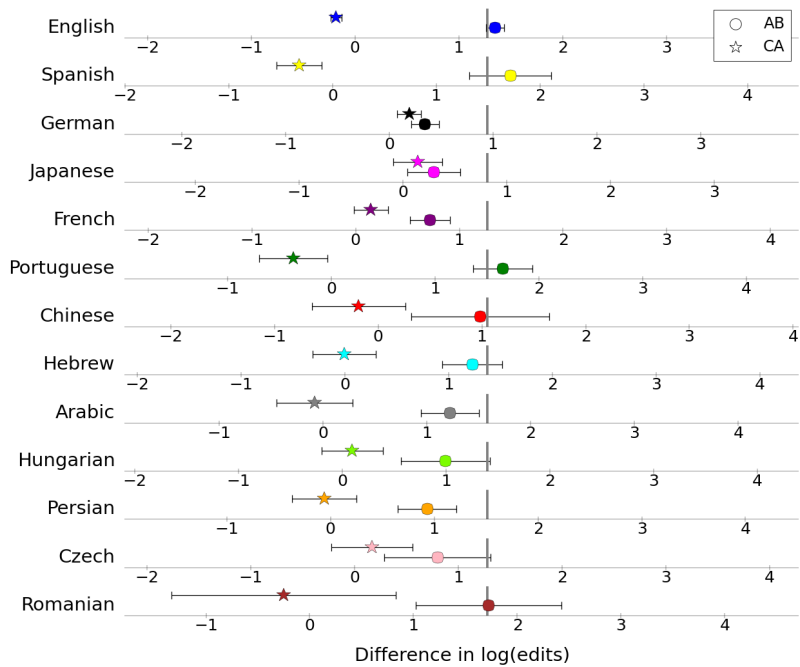

**Fig. S13.** Expected and observed difference in status between *A* and *B* (circles) and *C* and *A* (stars) for the *AB-CA* motif. Gray vertical lines show the expectation. Confidence intervals show 2.6 standard errors, corresponding to $p < 0.01$.

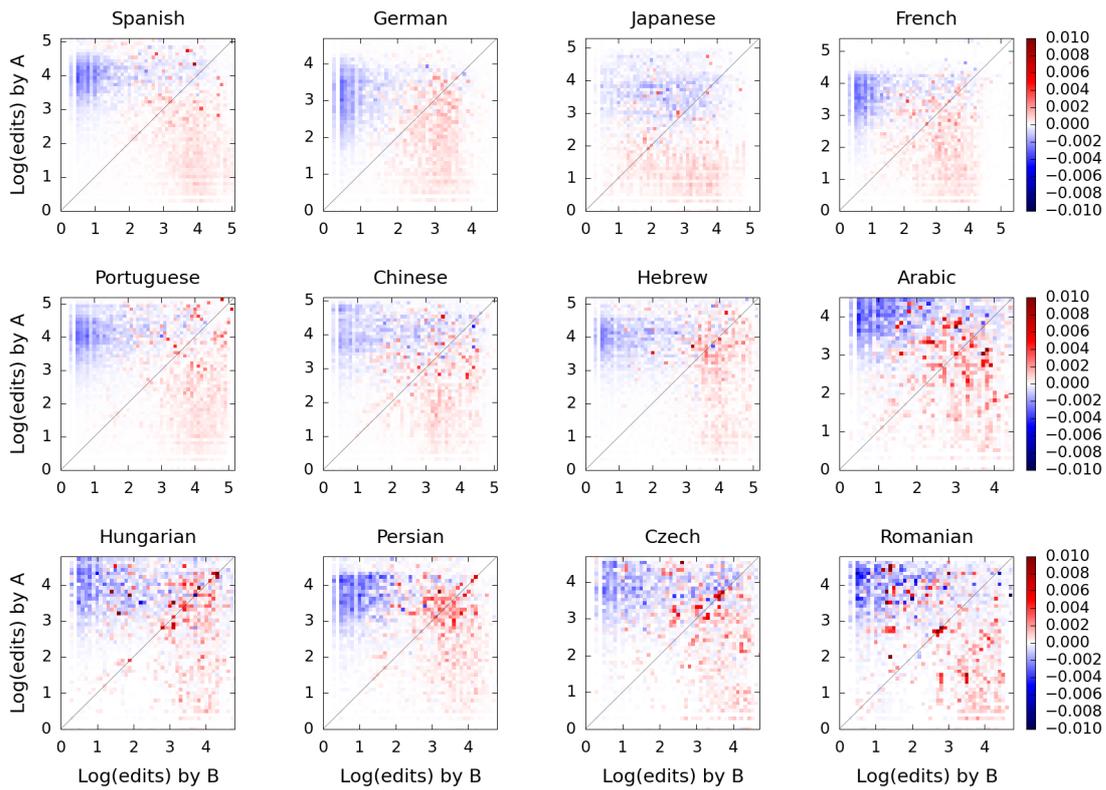

**Fig. S14.** Difference between the observed and expected distribution of status of *A* and *B* for the *AB-BA* motif.

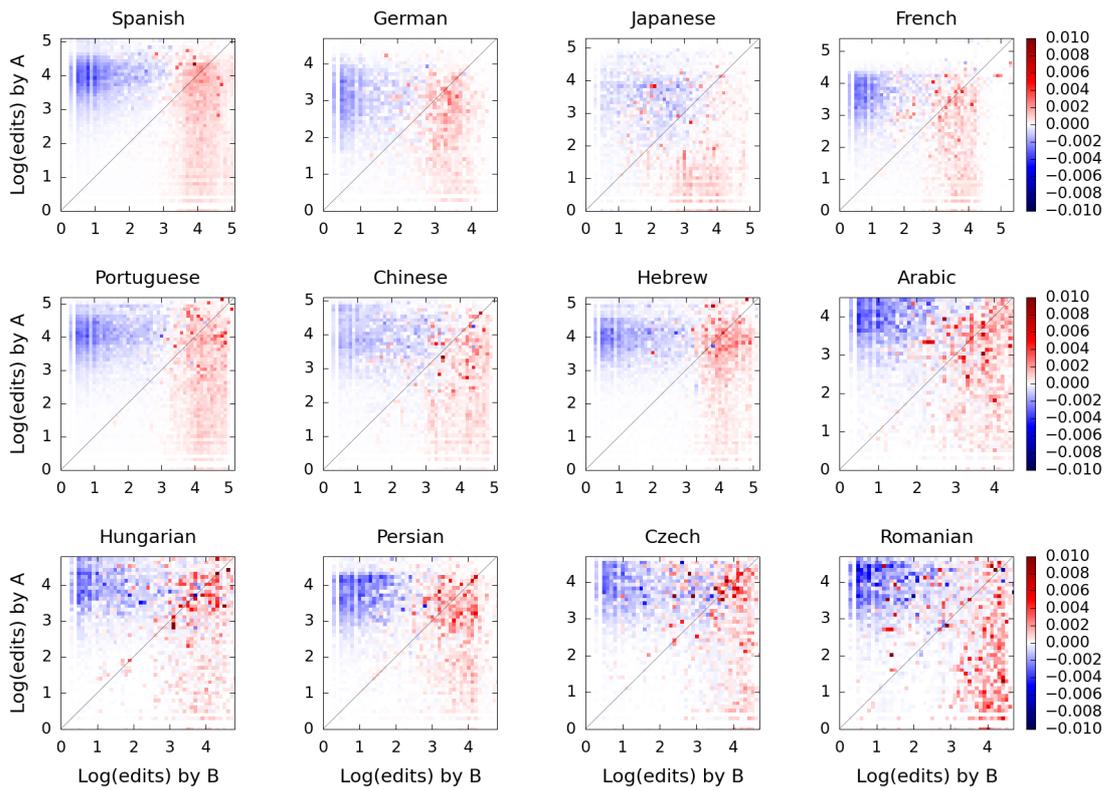

**Fig. S15.** Difference between the observed and expected distribution of status of *A* and *B* for the *AB-BC* motif.

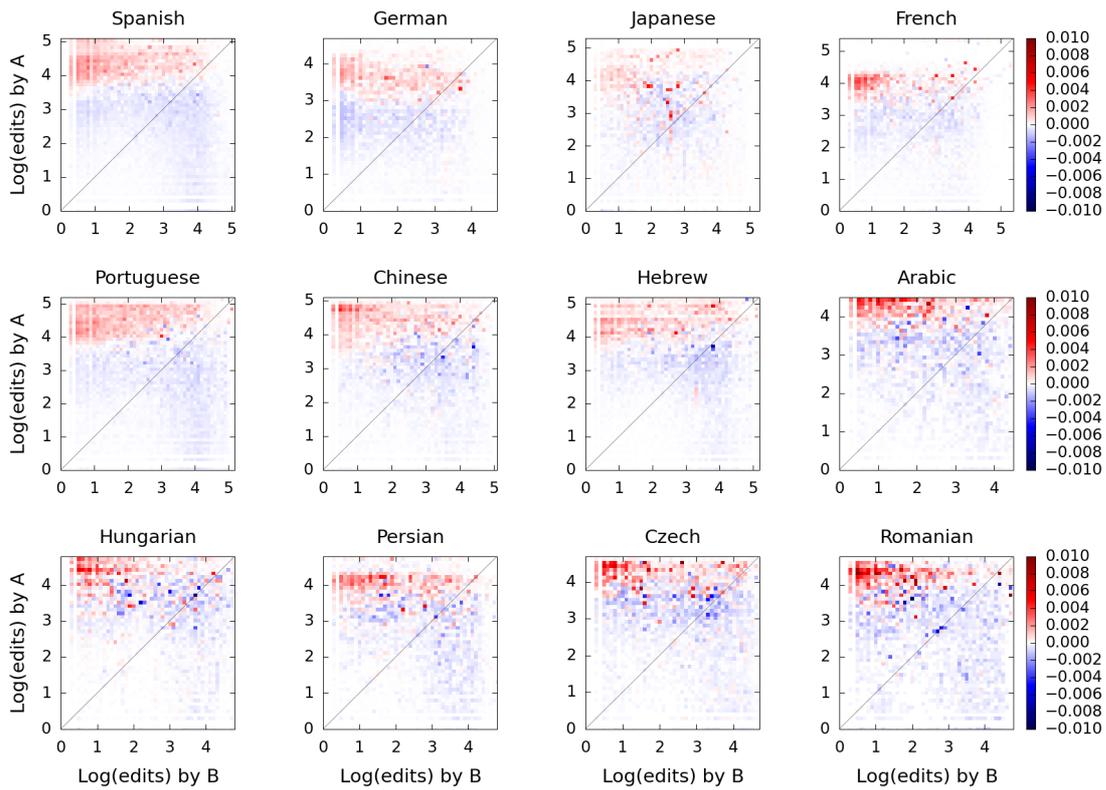

**Fig. S16.** Difference between the observed and expected distribution of status of *A* and *B* for the *AB-AC* motif.

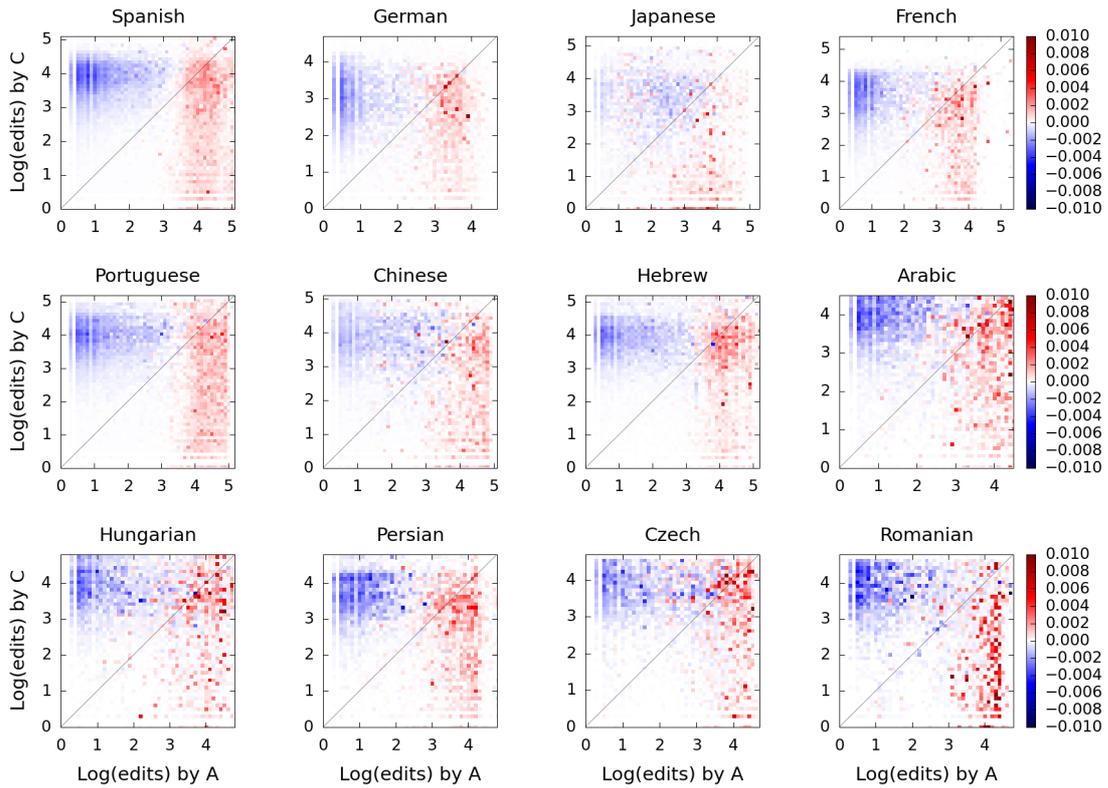

**Fig. S17.** Difference between the observed and expected distribution of status of *C* and *A* for the motifs associated with the *AB-CA* motif.

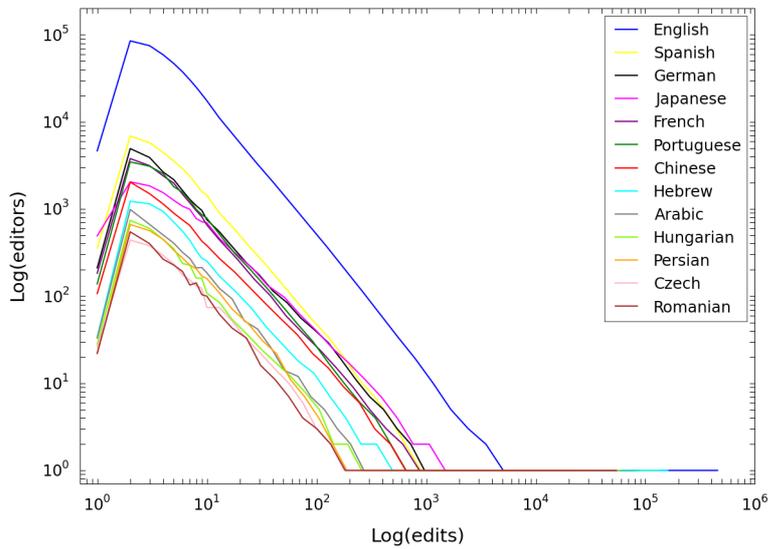

**Fig. S18.** Distribution of number of edits.